\documentclass[final,3p,onecolumn,authoryear]{elsarticle}

\usepackage[table]{xcolor}
\definecolor{myred}{rgb}{0.7961, 0.2627, 0.2039}  
\definecolor{mygreen}{rgb}{0.1294, 0.6039, 0.3294}  
\usepackage{booktabs}
\usepackage{amssymb}
\usepackage{amsmath}

\renewcommand{\thesection}{}

\usepackage[colorlinks=true, citecolor=blue]{hyperref}   
\usepackage{cleveref}   
\usepackage[labelfont=bf,labelsep=period]{caption}
\usepackage{booktabs} 
\usepackage{textcase}

\usepackage{xcolor}

\makeatletter
\def\ps@pprintTitle{%
  \let\@oddhead\@empty
  \let\@evenhead\@empty
  \let\@oddfoot\@empty
  \let\@evenfoot\@empty
}
\makeatother

\begin{document}

\hypersetup{linkcolor=blue, urlcolor=blue,citecolor=black}

\begin{frontmatter}

\title{\large{\textbf{Passive acoustic logic via topology-optimized  waveguides}}}
\makeatletter
\def\theaffn{\arabic{affn}}
\makeatother
\author[label1]{Ali Jafari}
\author[label1]{Mohamed Mousa}
\author[label1,label2]{Mostafa Nouh\corref{mycorrespondingauthor}}\cortext[mycorrespondingauthor]{Corresponding author}
\ead{mnouh@buffalo.edu}

\address[label1]{{Dept. of Mechanical and Aerospace Engineering, University at Buffalo (SUNY)}, {Buffalo}, {14260-4400}, {NY}, {USA}}

\address[label2]{{Dept. of Civil, Structural and Environmental Engineering, University at Buffalo (SUNY)}, {Buffalo}, {14260-4300}, {NY}, {USA}}  

\begin{abstract}
Growing energy demands of modern digital devices necessitate alternative, low-power computing mechanisms. When incident loads take the form of acoustic or vibrational waves, the ability to mechanically process information eliminates the need for transduction, paving the way for passive computing. Recent studies have proposed systems that learn and execute mechanical logic through buckling, bistability, and origami-inspired lattices. However, owing to the large timescales of shape morphing, such concepts suffer from slow operation or require active stimulation of adaptive materials. To address these limitations, we present a novel approach to mechanical logic, leveraging the rich dynamics of wave propagation in elastic structures. In lieu of traditional forward-design tools, such as band diagrams and transmission spectra, we employ a multi-faceted topology optimization approach, enabling us to identify candidate waveguide configurations within an extremely large design space. By incorporating voids within an otherwise uniform substrate, the optimized waveguides are able to precisely manipulate wave propagation paths, triggering desirable interferences of the scattered wavefield that culminate in energy localization at readouts corresponding to a given logic function. An experimental setup is used to demonstrate the efficacy of such logic gates and their resilience to non-uniform loading. By implementing these building blocks into a mechanical full adder, we demonstrate the scalable deployment of more sophisticated mechanical computing circuits, opening up new avenues in mechanical signal processing and physical computing.\\

\vspace{0.5em}
\noindent \textbf{Keywords} \\
mechanical intelligence, topology optimization, logic gates, wave-based computing
\end{abstract}


\end{frontmatter}

\section{Introduction}
\label{introduction}

Computing and signal processing lie at the heart of any modern technology. From mobile communications and autonomous vehicles to medical systems and civil infrastructure, and everything in between, they provide the tools and algorithms by which data can be analyzed and decoded. With the increasing demand and expanded utility of computing systems, requirements for pace, memory, data handling, and energy efficiency have significantly intensified in recent years. While electronic transistors and magnetic storage methods offer advantages in calculation speed and data density, their power consumption has exponentially increased \citep{Shankar_IEEE}, motivating the exploration of alternative computing paradigms. Of particular promise are systems that harness the energy inherently present in input signals, enabling them to be processed and interpreted in their same physical form. In this context, mechanical computing has recently emerged as a viable approach for materials and structures with inherent or embedded adaptive capabilities to intelligently respond to mechanical excitations such as vibrations, acoustic pressure, and thermal changes \citep{yasuda2021mechanical,he2024programmable,alu2025roadmap}. This notion builds on decades of research developments in the domain of smart architected materials (\textit{aka} metamaterials) which enable a programmable response through local property and micro-structural changes induced by external stimuli, including, but not limited to, magnetic \citep{pierce2020adaptive,moghaddaszadeh2023local}, thermal, ferroelectric \citep{hu20203d}, chemical \citep{hu2022releasing}, and light exposure \citep{patel2022photo}. The early wave of these efforts has culminated in programmable metamaterials with realizable logic \citep{wei2025programmable,fort2025thermoelastic}, neuromorphic intelligence \citep{moghaddaszadeh2024mechanical}, and implementable read-write operations \citep{li2022stimuli}, among others. Within the realm of elastodynamics, these systems leverage different physical phenomena to achieve low-power computing with minimal energy requirements \citep{kam2010design,chowdhury2011design}. Mechanical structures are generally more durable than electronic components and printed circuit boards, making them better suited for extreme operational environments where digital computing may be infeasible, such as ionizing radiation or significantly elevated temperatures \citep{elzouka2017high}. Unlike electronic systems, mechanical systems can leverage these harsh conditions to enable the desired functionality rather than being impaired by them \citep{dubvcek2024sensor}.

Driven by this vision, there has been a surging interest in mechanically-intelligent structures that adapt and respond to external loads in an informed manner \citep{mahboob2011interconnect,merkle2018mechanical,zhang2021hierarchical,drotman2021electronics}. Since Boolean and logical operators comprise the fundamental building blocks of digital algorithms, recent efforts have explored different ways to physically realize integrated mechanical circuits that mimic their electronic counterparts, offering a viable, energy-efficient, and seamlessly integrable alternative to conventional digital computing. However, to date, these efforts have predominantly relied on static or quasi-static deformations \citep{ansari20233d}, exploiting bi-stable mechanics \citep{zanaty2020reconfigurable,li2024reprogrammable,yang2025bistable} and buckling concepts \citep{hou2024mechanical,song2019additively,mei2023memory}, typically implemented in origami \citep{treml2018origami,yasuda2017origami} and kirigami-like structures \citep{korpas2021temperature,yasuda2020transition}, to physically realize logic and bit abstraction in mechanical form. Most, if not all, dynamic loads that impinge on deformable structures generate propagating waves within the elastic medium. Failing to capitalize on these waves, particularly given their versatility and faster travel times in finite substrates compared to static loading, represents a missed opportunity to revolutionize mechanical computing systems. Recent studies have begun to exploit such wave phenomena to realize acoustic and elastic wave-based logic operations, including implementations based on wave interference \citep{wang2019binary}, soft materials \citep{li2024realization}, and topological waveguides \citep{liu2025experimental}, demonstrating functionalities such as AND/OR gating and signal routing. Additionally, the ability of dispersive metamaterials to manipulate incident waves through attenuation bands \citep{al2017formation}, beaming effects \citep{ruzzene2003wave}, edge modes \citep{wang2015topological}, and localization \citep{al2023theory}, provides flexible strategies for wave energy manipulation. Understandably, these concepts have also been deployed to obtain new mechanical computing configurations in which guided elastoacoustic wave scattering represents the genesis of the computational methodology, unlocking new features such as frequency-selective \citep{dorin2024embodiment} and parallelized operations \citep{mousa2024parallel,mousa2025analog}.

\begin{figure*}[h!]
\centering
\includegraphics{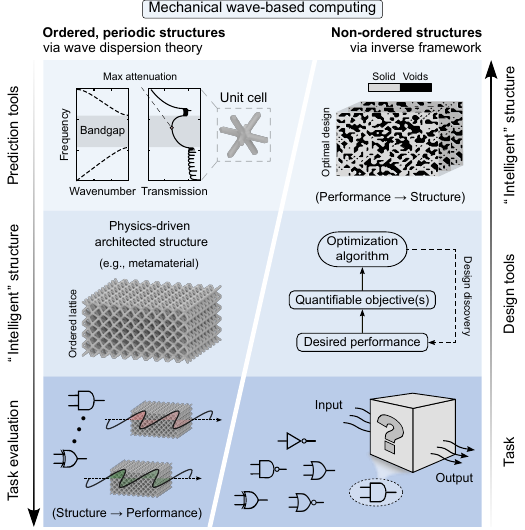}
\caption{Forward versus inverse design frameworks for mechanical wave-based computing. In a forward design approach, physics-driven tools such as dispersion diagrams and transmission spectra, among others, are used to realize candidate configurations which are typically ordered in shape and geometry and rely on human intuition. In contrast, an inverse design approach utilizes algorithmic (such as topology optimization) or data-driven tools to determine an optimal structure starting from a desired outcome, exploring a large design space in the process. This path typically produces non-ordered structures which, while effective in satisfying quantifiable objectives, are not necessarily intuitive.}
\label{Fig1}
\end{figure*}

When it comes to wave-based computing, the looming goal of achieving precise control over displacement/velocity fields runs into the challenges of complex wave scattering, including reflections, stability, and propagation losses associated with intricate geometries and heterogeneous material properties, typically found in highly-ordered and composite-like metamaterials. These hurdles often dictate design trade-offs which limit, if not kill, the intended functionality. This forward design approach is rooted in wave dispersion theory and physical norms, and relies on evaluating well-established tools (e.g., band structures, modal analysis, phase diagrams, transmissibility spectra, etc.) to identify suitable candidates, often foregoing a much larger design space which potentially subsumes non-intuitive solutions. In this work, we propose and explore a radically different approach in which we substitute this conventional approach with a blank-slate algorithm that is not constrained by predetermined patterns or periodicity requirements. Specifically, we employ a multi-faceted topology optimization (TO) approach \citep{bends2003topology,bendsoe1988generating,bendsoe1999material}, in order to identify optimal configurations which achieve complex desirable wave patterns under defined constraints and boundary conditions (see Figure~\ref{Fig1}). Since these configurations would algorithmically emerge from a robust inverse scheme, they no longer require actuation or external power in order to tune local components or unit cells. Therefore, in doing so, we seek to unlock new opportunities in \textit{passive} wave-based mechanical computing, which would be otherwise unattainable via conventional methods. 

A wide-ranging array of topology optimization methods have been developed and studied over the past three decades, often catering to different goals, from density-based subtractive methods \citep{zhou1991coc,rozvany2000simp} to evolutionary structural optimization approaches \citep{tanskanen2002evolutionary,xie1996evolutionary}. Recent applications of TO include reusable mechanisms for additive manufacturing \citep{tyburec2025modular}, improving strain distribution uniformity in sandwich flexible skins \citep{wei2025topology}, bistable flexure mechanisms \citep{wallin2021topology}, compact photonic devices \citep{he2022topology}, acoustic diodes \citep{bokhari2021topology}, and highly-absorbent broadband metamaterials \citep{feng2025topology}. In this work, we introduce TO as a foundational design mechanism for mechanical logic gates, i.e., optimized waveguides which manipulate incident waves to effectively execute logical operations. As will be demonstrated, the fundamental framework underpinning these structures provides substantial design flexibility and frees up the problem from physics-based restrictions which curtail operation or inadvertently diminish the available parameter space. The gate architecture, detailed in the next section, involves input and output deformable legs which are connected via a rectangular elastic substrate, initially set as a uniform solid region. Through finite element modeling, the TO-based process then proceeds to iteratively eliminate cells from this mesh such that the resulting wave propagation through the output legs adheres to the truth table of the gate of interest. The approach can be effectively applied to realize any fundamental logic gate, providing a powerful path toward integrated mechanical circuits that target task-specific outputs. Systems obtained from the presented optimization scheme are fabricated and experimentally tested, showing consistent and robust performance across different input scenarios, logic functions, and excitation amplitudes. Through the numerical-experimental framework, a fully passive mechanical adder circuit is then assembled from the respective constitutive gates, demonstrating the versatility of the individual gates and, more importantly, the system's immunity to unpredictable perturbations in the computational inputs which excite the waveguides in vibrational form. The study lays the foundation for higher-order multi-bit operations which form the basis of mechanical signal processors.

\section{Wave-based mechanical logic}
\label{theory}

The notion of in-material computing relies on the premise that a material's tuned response or adaptation to external loads can be intelligently correlated to such input, enabling information processing to be viewed as an embedded material property. When such loads are mechanical in nature, structural deformations, elastic strains, and kinematic stability can be exploited to process external stimuli, enabling arithmetic and combinational logic to take shape within a material \citep{el2022mechanical}. Achieving this via wave propagation rather than quasi-static changes of state drastically improves computational pace and latency but comes with its fair share of challenges. Our focus here is on a class of mechanical logic gates in which incident waves representing the binary inputs of a Boolean operation are made to propagate in a manner that localizes downstream energy at a readout location consistent with the operation's outcome \citep{fort2025thermoelastic}. The convoluted nature of elastoacoustic waves, ranging from complex scattering and modal interactions to the coupled effects of material and geometry, can render the forward-design approach of such systems exhaustive and often futile. Alternatively, topology optimization provides a powerful tool by which a large design space can be explored to yield intricate designs that achieve this goal while tackling the aforementioned challenges. Such designs would remain practically unrealizable through intuition alone. 

\subsection{Gate architecture}
\label{gate_architecture}
\begin{figure*}[h!]
\centering
\includegraphics[width=\linewidth]{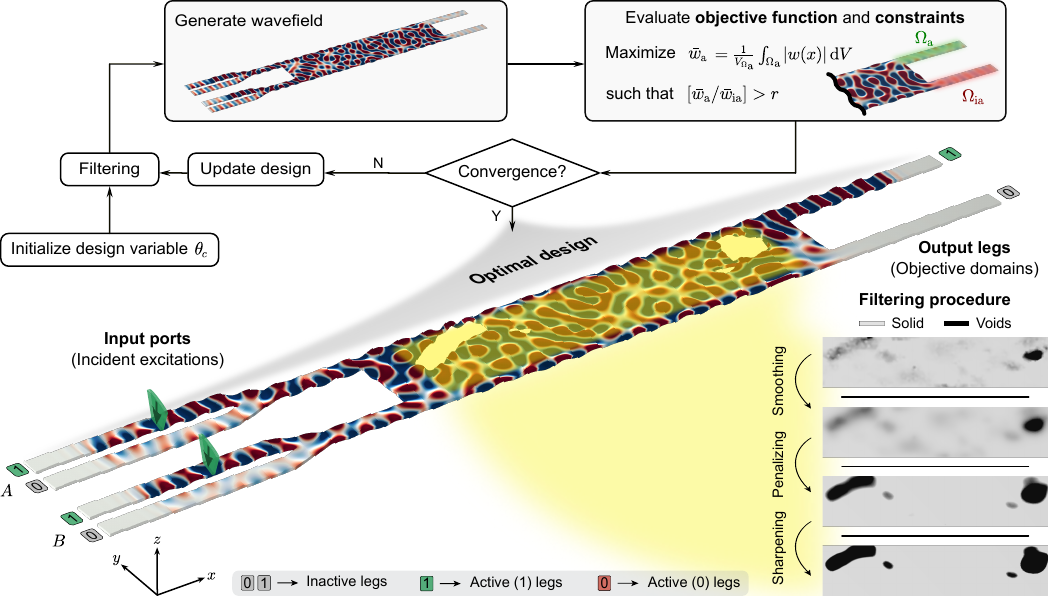}
\caption{Flowchart representation of the topology optimization algorithm and the associated filtering procedure of a wave-based mechanical logic gate. The optimization framework begins by initializing the design variable $\theta_c$, followed by filtering and wavefield generation. The optimizer maximizes the objective function, $\bar{w}_{\mathrm{a}}$, subject to the constraint $\bar{w}_{\mathrm{a}} / \bar{w}_{\mathrm{ia}} > r$. The waveguide design is updated iteratively until convergence is achieved. The central schematic shows an optimal design resulting from this process. The filtering procedure inset details the post-processing of the material density field to enforce a binary material distribution for fabrication purposes.}
\label{Fig2}
\end{figure*}

An electronic logic gate, such as AND or OR, is comprised of two inputs and a single output. The state of each input or output is determined by the level of electrical current flowing through it, with a low current representing a (0) state and a high current denoting a (1) state, typically $4$ and $20$ mA, respectively. The non-zero low threshold accounts for power loss such that levels below $4$ mA can indicate a dead signal or an inactive state. Contrary to digital systems where a signal varies virtually via discrete values of both time and amplitude, physical computing platforms (of which mechanical systems is a subset) are fundamentally analog in nature and rely on the continuous variation of physical parameters to correlate an output of a computational task to its input. Binary logic in these systems, therefore, emerges through interpretation mechanisms such as voltage, intensity, or displacement thresholds. We design the mechanical analog of a logic gate in the form of an elastic waveguide which receives input through vibroacoustic excitations. Consistent with its electronic counterpart, the mechanical gate, which is depicted in Figure~\ref{Fig2}, incorporates two input ports, each represented by a two-prong fork. During operation, only one of these two prongs, henceforth referred to as ``legs'', is activated by virtue of incident waves that propagate within it. An input is determined to have a state of (1) when waves flow through the upper leg and a state of (0) when they flow through the lower leg of that input. The input forks feed into a rectangular elastic material, which serves as an optimization domain, where the input waves scatter and spatially interfere. The gate culminates with two output legs that branch out of the optimization domain, as shown in the same figure. Similar to the input ports, a dominant wave presence in the upper output leg denotes an outcome of (1), whereas dominant waves in the lower leg indicate a (0) outcome. As such, an input or output can only have one active leg at a time. The overarching goal is to optimize the design of the middle region connecting the gate's output with its inputs.

To realize the intended logic operation, the optimizer seeks to fine-tune the material composition of the optimized domain such that the overall waveguide functions as the desired gate. For this goal to be achieved, the optimized design needs to yield all the different wave propagation patterns commensurate with the different scenarios (generally four unique scenarios) of a logic gate's truth table. We achieve this by subtractive design, i.e., removing a minimal amount of superfluous material so that the resultant structure functions as needed. The aforementioned approach provides a tangible path toward practical realization since removed material can be replicated by carving out small holes, for example, via waterjet or laser cutting, creating precise voids within the elastic substrate.

Consider the mechanical AND gate, displayed in Figure~\ref{Fig2}, with the two input ports, $A$ and $B$. In the example shown, the gate is subject to two (1) inputs, i.e., $A = B = 1$, with an expected outcome of (1), i.e., $A \land B = 1$. As a result, the optimal structural configuration is one that effectively redirects energy from these two incident waves, as they travel through the middle region, to the upper output leg on the far right side of the gate, indicating an output state of (1). In practice, however, a small amount of wave energy is inevitably bound to leak into the inactive (lower in this case) output leg. To quantify and evaluate this goal, we utilize elastic deformation measurements within the two output legs. As such, a successful topology is defined as one where the out-of-plane displacement within the active output leg, $\bar{w}_{\mathrm{a}}$, proves to be significantly larger than that of the inactive one, $\bar{w}_{\mathrm{ia}}$, at steady-state conditions. Both displacements are averaged over the volumes of their corresponding output leg, $V_{\Omega}$. In the previous variables, the subscripts ``$\mathrm{a}$'' and ``$\mathrm{ia}$'' denote the active and inactive output legs, respectively. As an example, in the AND gate shown in Figure~\ref{Fig2}, $\bar{w}_{\mathrm{a}}$ and $\bar{w}_{\mathrm{ia}}$ denote the averaged displacement in the upper and lower output legs, respectively. To formally cast this problem, the primary objective of the optimizer is therefore to maximize the average displacement within the active leg, $\bar{w}_{\mathrm{a}}$, with the added constraint that a final gate design is only deemed acceptable if the ratio of the active-to-inactive displacements exceeds a preset threshold $r$, i.e., $[\bar{w}_{\mathrm{a}}/\bar{w}_{\mathrm{ia}}] > r$ for a given set of inputs. As such, for a single topology to be successful in executing the entire truth table for the AND gate, the goal is to maximize the aggregate active leg displacements, $\Sigma_{i=1}^I \bar{w}_{\mathrm{a},i}$, where $I=4$ represents the number of possible input scenarios, such that $i= 1$, $2$, $3$, or $4$ defines the different truth table rows $[A,B]=[0,0]$, $[1,0]$, $[0,1]$, and $[1,1]$, respectively, while ensuring the constraint $[\bar{w}_{\mathrm{a},i}/\bar{w}_{\mathrm{ia},i}] > r$ is simultaneously satisfied for all $I=4$ scenarios.

In the remainder of this section, we lay out the foundations of the TO algorithm, detail how a single, unified design that satisfies the aforementioned objectives can be obtained for a given logic gate, and evaluate the performance of the topology-optimized gates.

\subsection{Topology optimization}
\label{TO}
Topology optimization is an inverse design technique that determines an efficient material and/or geometric layout that best satisfies a design objective and a set of constraints. In this work, a density-based TO algorithm is employed, where an artificial material density variable, $\theta_c$, acts as the primary control parameter for the structural design. This parameter typically ranges from $0$ to $1$, with $\theta_c = 0$ representing voids (no material), and $0 < \theta_c \leq 1$ denoting the presence of material whose stiffness and density progressively increase as $\theta_c$ approaches $1$. As such, $\theta_c$ effectively governs the local stiffness and density distribution within the design domain. While the process starts from a graded middle region with an initial material composition $\theta_0$ that spans the $\theta_c$ range, the final configuration of the mechanical logic gate will eventually be binary, consisting of an aluminum substrate ($\theta_c = 1$) that contains voids ($\theta_c = 0$).

To start the design process, we mathematically define the TO problem as follows:
\begin{equation}
\label{eq:TO}
\begin{aligned}
\textrm{max} \quad &f_0(\theta_c)\\
\textrm{subject to} \quad &f_i(\theta_c) < 0, \quad i=1, \dots ,I \\
&\theta_c \in [0,1]
\end{aligned}
\end{equation}
where $f_0(\theta_c)$ denotes the objective function with $I$ number of constraints, $f_i(\theta_c)$. While equation~(\ref{eq:TO}) defines a generalized framework for TO, in this problem the maximizable function $f_0(\theta_c)$ is the averaged displacement within the gate's output leg, $\bar{w}_{\mathrm{a},i}$, and consequently, $f_i(\theta_c)$ is represented by $(r - [\bar{w}_{\mathrm{a},i}/\bar{w}_{\mathrm{ia},i}])$, where $r=1.5$ is deemed an adequate threshold throughout the study. We employ the method of moving asymptotes (MMA), which addresses intricate non-linear programming challenges by formulating and resolving a sequence of strictly convex subproblems. These subproblems employ \textit{moving asymptotes} to adaptively regulate the approximation, enhancing both optimization stability and convergence speed \citep{svanberg1987method}.

While the optimization initially allows the continuous design variable $\theta_c$ to take any values from $0$ to $1$, the intermediate values correspond to hypothetical material properties which are not necessarily physically realizable, and regardless, would not be utilized in a binary (solid-voids) design. Consequently, a series of filters and penalization techniques are imposed, as depicted in the bottom right corner of Figure~\ref{Fig2}, to ensure the final binary design retains similar functionality to that of the originally-obtained optimal configuration that encompasses a continuous $\theta_c$ distribution. Following the definition of objective and constraint functions within a finite element (FE) model, $\theta_c$ is initialized with a value of $1$ for all domain elements within the middle region of the gate. The optimizer then proceeds to randomly adjust $\theta_c$ values, effectively redistributing material within the design domain, and progressively generates patterns that satisfy the problem requirements. However, these patterns typically resemble mesh-dependent composites with minimal features, i.e., infinitesimal elements that would result in infeasible designs that are practically unmanufacturable. Therefore, a Helmholtz partial differential equation (PDE) is used as an efficient filter to regularize the design variables and smooth out the material distribution. This avoids the creation of checkerboard-like patterns that can be physically impossible to realize, and rather encourages the optimizer to search for continuous and feasible designs. This filter can be described as follows \citep{lazarov2011filters}
\begin{align}
\theta_f &= \theta_c + R_{\mathrm{min}}^2 \nabla^2 \theta_f
\end{align}
The filtering process, realized by solving this Helmholtz PDE, generates a smoothed distribution variable $\theta_f$ from the control variable $\theta_c$, such that all features smaller than a designated minimum length scale, $R_\textrm{min}$, are eliminated. The downside of this, however, is that the Helmholtz filter gives rise to a significant grayscale design which doesn't have a clear physical interpretation, i.e., with many elements in the design domain that have $\theta_f$ values between $0$ and $1$. To mitigate this, a projection step is implemented to reduce the extent of the grayscale region based on the following hyperbolic tangent function \citep{wang2011projection}
\begin{align}
\theta &= \frac{\tanh\big(\beta(\theta_f - \theta_\beta)\big) + \tanh\big(\beta\theta_\beta\big)}{\tanh\big(\beta(1 - \theta_\beta)\big) + \tanh\big(\beta\theta_\beta\big)}
\end{align}
where $\theta$ is the output material volume factor, and the steepness parameters, $\beta$ (projection slope) and $\theta_\beta$ (projection point), govern the projection function. The goal is to apply a smooth step function that projects intermediate densities towards $0$ or $1$, with the steepness of the projection being proportional to the value of $\beta$. Specifically, low $\beta$ values facilitate gradual transition toward binarization, whereas high $\beta$ values enforce nearly binary distributions. A continuation strategy is typically required, wherein $\beta$ is gradually increased throughout the optimization process to maintain stable convergence while progressively rendering the design more discrete. The projection function generates a nearly binary physical density field; however, it does not inherently specify the incorporation of material properties into the governing equations. To guarantee physically relevant penalization of intermediate densities and to uphold numerical stability, the solid isotropic material with penalization (SIMP) interpolation is subsequently employed on the projected field. The penalized material field is characterized by
\begin{align}
\theta_p &= \theta_{\mathrm{min}} + (1 - \theta_{\mathrm{min}})\theta^{p_\mathrm{SIMP}}
\end{align}
where $\theta_{\mathrm{min}}$ is the minimum volume fraction allowed in the model and ${p_\mathrm{SIMP}}$ is the SIMP penalization exponent (typically ${p_\mathrm{SIMP}}=3$ for compliance minimization problems). This prevents the stiffness from being zero anywhere in the design domain for numerical reasons. Unlike the projection function, which enforces geometric discreteness, SIMP penalizes intermediate densities through nonlinear constitutive interpolation, rendering them suboptimal in the optimization process \citep{guest2004achieving}. Once $\theta$ and $\theta_p$ are calculated, they are directly applied to the stiffness, $E=\theta_{p}E_{Al}$, and density, $\rho=\theta\rho_{Al}$, of the domain elements within the optimized region, where $E_{Al}$ and $\rho_{Al}$ represent the elastic modulus and density of aluminum, respectively.

With the updated control parameter $\theta_c$ and its variations calculated after the first iteration of optimization, the wave pattern is numerically re-evaluated using the updated stiffness and density, and the objective and constraint values are calculated accordingly. As shown in Figure~\ref{Fig2}, this entire process is repeated until convergence is reached. In the first round of optimization, a low $\beta$ value is utilized to catalyze the optimizer and enable it to quickly find a solution, generating a design that satisfies the problem requirements. However, this design would be far from binary due to the low steepness parameter used. Following this, $\beta$ is increased gradually, and additional optimization rounds are performed, with the outcome generated by each round used to initiate the following one. This approach significantly speeds up the optimization process and improves its convergence. While the final design generated by the optimizer is almost perfectly binary, a verification step is necessary to test a completely binary design after applying a hard filter on the final solution to generate a $\theta$ distribution with values that are strictly $0$ or $1$. We refer to the latter as the binary output material index, $\theta_b$. The optimization is deemed successful if a mechanical gate utilizing the verified design adheres to the truth table outcomes with a displacement ratio between the active and inactive legs that exceeds the preset threshold.

\begin{figure*}[h!]
\centering
\includegraphics[width=\linewidth]{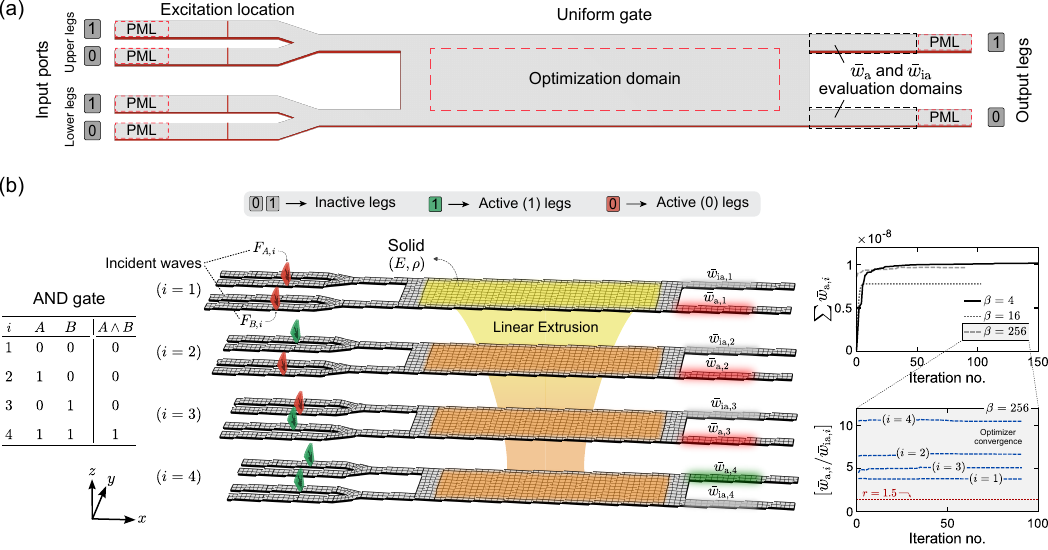}
\caption{(a) Uniform gate configuration (pre-optimization), showing all input and output legs, the optimization domain, and incident wave locations. Output regions where the averaged out-of-plane displacements, $\bar{w}_{\mathrm{a}}$ and $\bar{w}_{\mathrm{ia}}$, are evaluated in each iteration of the optimization process are marked by black dashed boxes. (b) Schematic diagram outlining the process of obtaining a unified gate design per logic operation through an example AND gate. For better visualization, the displayed mesh comprises larger elements than those used in the simulations. The truth table of the AND gate is shown on the left for all four possible input scenarios, $i$. Excitation forces at the input ports $A$ and $B$ are marked by ${F}_{A,i}$ and ${F}_{B,i}$, respectively. All active (1) legs are highlighted in green, active (0) legs are highlighted in red, and inactive legs are left gray. The optimizer maximizes the aggregate average displacement of the active legs across all four input scenarios, $\Sigma_{i=1}^4 \bar{w}_{\mathrm{a},i}$, with four constraints concurrently imposed ensuring that the ratio of active-to-inactive output displacements exceeds the preset threshold in each of the four cases separately, i.e., $[\bar{w}_{\mathrm{a},i}/\bar{w}_{\mathrm{ia},i}] > 1.5$. To guarantee a unified waveguide design for all $i$ values, the optimized geometry from the ($i=1$) case is replicated across the remaining three cases for each iteration. The evolution of the objective function for select $\beta$ values is illustrated in the top plot on the far right side, whereas the variation of the four concurrent constraints throughout the iterative process for $\beta=256$ step is shown on the bottom plot. While the optimizer is set to run up to $150$ iterations for each $\beta$ step, a step concludes earlier if convergence is reached, i.e., when changes in the objective fall within a tolerance of $10^{-6}$, as captured by the two plots.}
\label{Fig3}
\end{figure*}

\subsection{Unified gate design per logic operation}
\label{unified}

Obtaining a mechanical structure that satisfies a single objective, corresponding to one of the four truth table scenarios, $i$, following the process outlined above is relatively straightforward. However, the challenge resides in integrating all four cases within a unified gate design. To achieve this unified design, a comprehensive FE-based model is first constructed using the Structural Mechanics and Optimization modules within \textsc{COMSOL} Multiphysics\textsuperscript{®}. In all forthcoming time and frequency domain simulation results, gates are excited at the left edge via input flexural waves carrying a single operational frequency of $92$ kHz. While the frequency choice has no bearing on the general design concept, this value was found to be adequate in terms of activating higher-order wave modes with wavelengths that are sufficiently smaller than the overall size of the optimized domain, thereby giving the optimizer flexibility to find designs that effectively manipulate incident waves within the allotted space. We also note that the overall gate size and the choice of out-of-plane displacements are best suited for experimental testing, allowing a 1D Scanning Laser Doppler Vibrometer (SLDV) system, which only measures velocities normal to the laser, to provide a clear visualization of the wavefield. However, the framework presented here can be readily extended to different modes and length scales.

The unified gate design process begins with defining the gate geometry, including the two input legs, the central design domain, and the two output legs, with a universal thickness of $1$ mm, as depicted in Figure~\ref{Fig3}(a). This geometry is duplicated to create $I=4$ instances, each representing one of the logic gate scenarios, $i$, of its truth table, such that $i = 1, \dots, I$, as defined earlier. While we only seek one unified physical design, these duplicates represent hypothetical gates that enable us to concurrently optimize across different input scenarios. In other words, the primary difference among these instances is where the edge excitations are imposed at the input legs. For example, for the case when $i = 2$, i.e., $[A,B] = [1,0]$, vibrational waves enter the system at the upper leg of the first input and the lower leg of the second one, following the general configuration described earlier. The model is augmented with perfectly matched layers, marked as PML on Figure~\ref{Fig3}(a), to mitigate edge reflections. The mesh is subsequently defined with element sizes fine enough to capture the intricate details of wave propagation. The maximum element size is set to about one-twentieth of the wavelength at the operating frequency, while elements near boundaries and around small voids introduced by the optimizer are further refined to ensure accurate resolution of local wave scattering phenomena.

The next step is to set up the optimization problem, starting with the objective function $f_0(\theta_c)$ and associated constraints $f_i(\theta_c)$. Out-of-plane displacement domain probes are defined over the output legs to monitor the average displacement amplitudes of the expected active ($\bar{w}_{\mathrm{a},i}$) and inactive ($\bar{w}_{\mathrm{ia},i}$) legs across all four instances. This is illustrated in Figure~\ref{Fig3}(b), showing the configuration and design details of an example AND gate, where input and output legs exhibiting an active (1) state are highlighted in green, while those exhibiting an active (0) state are highlighted in red. The objective function to be maximized is formulated as the sum of the average displacements of the active output legs ($\Sigma_{i=1}^I \bar{w}_{\mathrm{a},i}$), while $I$ independent constraints are imposed such that the ratio between the average displacements of the active and inactive legs in each instance remains above the preset threshold of $r=1.5$, i.e., $[\bar{w}_{\mathrm{a},i}/\bar{w}_{\mathrm{ia},i}] > r$ for $i = 1, \dots, I$. Since we practically seek a single gate design that maximizes the aggregate of the four objectives and simultaneously satisfies the four constraints, the design domain for topology optimization is defined as that of the first instance ($i=1$) only, marked by the yellow-highlighted region in Figure~\ref{Fig3}(b). As a result, the material distribution iteratively developed in this first instance must be consistently reflected across the other three (orange-highlighted) domains. This linkage is achieved through the linear extrusion feature in \textsc{COMSOL\textsuperscript{®}}, which attains the stiffness $E_1(x,y)$ and density $\rho_1(x,y)$ fields within the first ($i=1$) design domain, and effectively duplicates them across the remaining instances, i.e., $E_{2-4}(x,y)$ and $\rho_{2-4}(x,y)$.

Following the design process, the FE model is run to solve for the wavefield of the gate's four instances, with the inputs properly set up for each case. Then, the objective function and the four constraints are re-evaluated, and the density map of the first instance's design domain will be updated by the optimizer and transferred to the other three domains via the linear extrusion. The projection parameter $\beta$ is incrementally increased from $2$ to $256$, doubling at each interval to maintain optimization stability and convergence, with up to $150$ iterations being performed for each value of $\beta$. This approach ensures the convergence of all four instances toward a single, unified structural design that consistently realizes the desired logical behavior for all input conditions. Once this process is complete, a single topology (i.e., material distribution) for the optimization domain is obtained, which is capable of directing the incident waves to the correct output leg regardless of the input scenario, $i$, that is applied.Finally, it is worth noting that topology optimization is a non-convex problem, and different choices of initialization may lead to different local optima  (Additional details regarding non-convexity and initialization effects are provided in Appendix~\ref{non-convex}). In this work, an unbiased uniform initial condition of $\theta_0=1$, corresponding to a fully solid design domain, is adopted as the baseline starting point for the optimization. Also, a volume constraint is imposed such that the optimized voids are restricted to occupy no more than $20\%$ of the optimization domain ($V_{\mathrm{opt}}\leq0.2$). This choice provides a controlled and manufacturable design space while allowing the optimizer sufficient freedom to realize the desired logic functionality.

\begin{figure*}[h!]
\centering
\includegraphics[width=\linewidth]{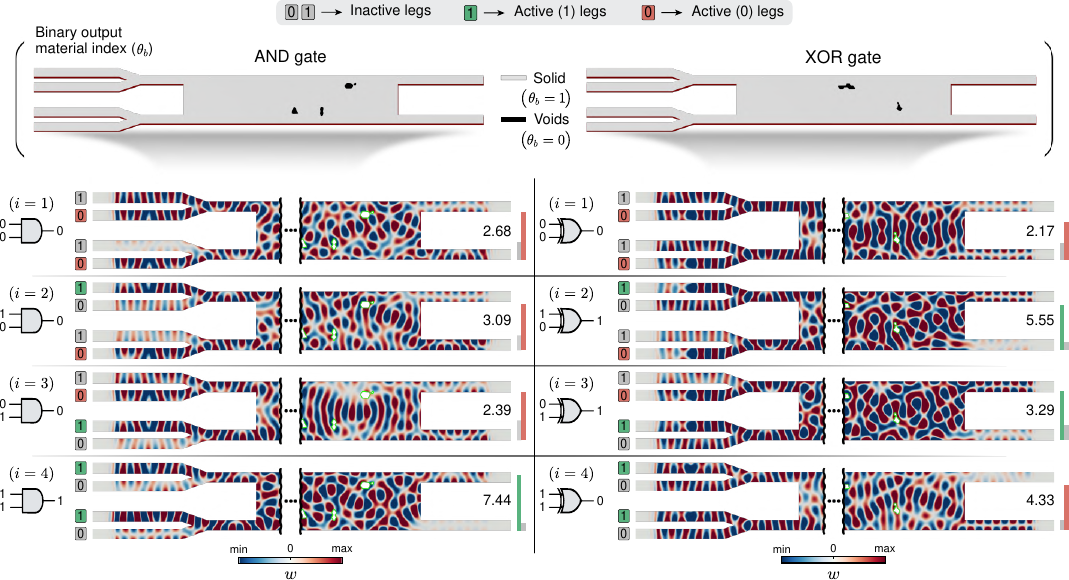}
\caption{Performance of the optimized wave-based mechanical logic gates. The top panel displays the final optimized gate designs for AND (left) and XOR (right) gates, whose binary output material index, $\theta_b$, indicates the fraction of solid material remaining after spatial placement of voids within the optimization domain. The lower panels show the displacement wavefields for both gates across all four cases of the truth table, i.e., $i=1$ through $4$. The bar graphs adjacent to the output legs of each wavefield represent the average displacement magnitudes in the active, $\bar{w}_{\mathrm{a},i}$, and inactive, $\bar{w}_{\mathrm{ia},i}$, legs, with green bars denoting (1) outputs, red bars denoting (0) outputs, and gray bars denoting inactive legs.}
\label{Fig4}
\end{figure*}


\subsection{Performance of topology-optimized gates}
\label{numerical_results}

Following the process outlined in the previous two sections, any of the fundamental logic operations can be conducted in mechanical wave-based form. The upper row of Figure~\ref{Fig4} displays the final optimized density maps for two example gates, namely AND and XOR. Black regions within the middle portion of the gate (optimized domain) indicate the voids needed to achieve the intended logical behavior. The rest of the domain represents solid aluminum regions. The remainder of Figure~\ref{Fig4} shows the wavefields for each of the four input scenarios associated with each gate, represented by surface contours of the out-of-plane displacement, $w$. The wavefields clearly demonstrate the ability of the mechanical gates to effectively realize the correct outcome of the intended logic operation for each and every input combination. Consider the ($i=4$) case with the inputs $[A,B]=[1,1]$ and an expected outcome of (1) for the AND gate shown in the bottom left of Figure~\ref{Fig4}. Waves are admitted through the upper leg of each input port, and the gate achieves a ratio of $[\bar{w}_{\mathrm{a},4}/\bar{w}_{\mathrm{ia},4}] = 7.44$ between the displacements of its active (upper) and inactive (lower) output legs, successfully producing a logical output of (1) and satisfying the displacement ratio constraint by surpassing the threshold value of $1.5$. The ($i=4$) case of the XOR gate exhibits a similar performance, where the same set of inputs is expected to yield a logical outcome of (0). As indicated in the lower right plot of Figure~\ref{Fig4}, this is achieved with an output displacement ratio of $4.33$. The remaining wavefields show a consistent performance, further confirming the ability of both gates to function as needed, and demonstrating such reliable performance with clear separation of logic states and high wave transmission across all input scenarios. We also note that the output ratios attained for the AND gate in Figure~\ref{Fig4} are slightly lower than those observed in the iteration outcomes in Figure~\ref{Fig3}(b). This reduction in performance is due to the hard filter applied to the output material factor, $\theta$, to obtain a strictly binary design, $\theta_b$.

The approach introduced here essentially enables the design of mechanical structures that represent any fundamental logic gate. Importantly, the initial uniform design's symmetry along the horizontal $x$-axis allows dual functionality between gate pairs that represent logical conjunction and disjunction. For example, flipping the AND gate vertically transforms it into an OR gate. The upper and lower legs of each input port always correspond to logical states of (1) and (0), respectively. As such, when an AND gate is flipped vertically, this labeling is reversed, such that the (0) leg becomes the (1) leg and vice versa. As a result, the ($i=1$) and ($i=4$) scenarios are interchanged, whereas the ($i=2$) and ($i=3$) cases remain unchanged, thereby complying with the expected outcomes of an OR gate (see Appendix~\ref{AND–to–OR} for more details).

The mechanical gates shown in Figure~\ref{Fig4} are evaluated through frequency domain simulations, effectively depicting the performance of each case at steady state. While all the shown cases demonstrate consistent functionality, certain input configurations appear to result in a comparable amount of wave energy in both upper and lower legs of each input port, suggesting a violation of the requirement that only one leg at a time can be activated in each input port. This is, for example, highly pronounced in the ($i=1$) case of both the AND and XOR gates, corresponding to $[A,B]=[0,0]$, making it practically difficult to infer the gate's inputs, i.e., the values of $A$ and $B$, from these wavefields alone. This situation is in fact attributed to wave reflections and backscattering from within the gate structure itself, which at times leak into the inactive input legs, albeit with no bearing or detrimental effect on the gate's response or its ability to obtain the correct computational outcome. The intensity of these backward-propagating waves varies between the different cases, with some configurations exhibiting stronger backfilled inactive legs than others. To confirm this, a full-fledged time domain analysis is performed to track the propagation of waves from their incidence at the active input legs, their subsequent scattering through the middle topology-optimized domain with the different voids, up until their arrival at the gate's output legs. 

\begin{figure*}[h!]
\centering
\includegraphics[width=\linewidth]{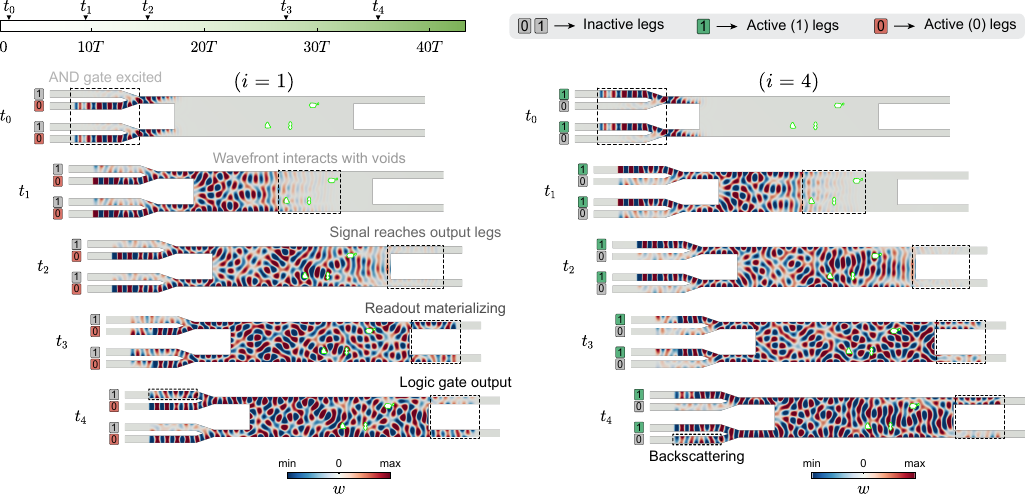}
\caption{Time evolution of the wave-based AND gate subject to two sets of inputs, $i=1$ and $4$. Wavefields are shown for both cases at five snapshots,  $t_0$, $t_1$, $t_2$, $t_3$, and $t_4$, where $t_0$ is shortly after input excitations are applied and $t_4$ shows the gate's response as the correct readout manifests itself in the output legs. Significant milestones are indicated using the dashed boxes. The time domain analysis demonstrates the progression of the gate's performance in both cases, and captures the emergence of reflected waves over time, which backfill some of the inactive input legs. The eventual wavefields show close agreement with the steady state (frequency domain) results of Figure~\ref{Fig4}, and confirm the negligible effect of backscattering on the gate's outputs across all input scenarios.}
\label{Fig5}
\end{figure*}

The results, depicted in Figure~\ref{Fig5}, show the transient wavefield evolution for the ($i=1$) and ($i=4$) cases, corresponding to $[A,B]=[0,0]$ and $[1,1]$, respectively, of the AND gate. Wavefield snapshots are displayed at five time instants, $t_0$, $t_1$, $t_2$, $t_3$, and $t_4$, corresponding to $0.38\times10^{-3}$, $0.1$, $0.15$, $0.27$, and $0.39$ ms, respectively, starting from shortly after the initial excitation. A corresponding timeline is indicated at the top of Figure~\ref{Fig5} as a function of the gate's oscillation period, $T$, which is approximately $10.9$ $\mu$s for the chosen operational frequency. Upon applying the inputs, the waves propagate through the input legs and into the optimized domain with little to no reflections initially (snapshots at times $t_0$ and $t_1$). This can be observed from the near-identical wavefields (vertically flipped due to mirrored input locations) of both cases up until time $t_1$. Following this, the waves interact with and reflect from the voids within the optimized domains on their way to the output legs, leading to observable backscattering toward the input ports, reshaping the incident wavefield pattern (snapshots at times $t_2$ and $t_3$). During these intermediate stages, the structure remains in a transient state, and the outputs are not definitive. After nearly $35$ oscillation cycles, the gate's response approaches a steady state, and the wavefields become consistent with the patterns demonstrated by the frequency domain results shown in Figure~\ref{Fig4}, as revealed by the results at time $t_4$. At this final stage, each gate successfully produces the correct logical outputs. However, minor back-propagation persists through the inactive (upper) leg of the $A$ and $B$ input in the ($i=1$) case as well as the inactive (lower) leg of the $B$ input in the ($i=4$) case, both of which were not originally excited. While this issue could be mitigated by introducing additional constraints to suppress backward waves in the optimization problem, such an approach would significantly increase computational cost and may compromise optimization performance. Importantly, this phenomenon does not affect the performance of the forward path, as further demonstrated in the mechanical full adder example presented later. The results presented here also provide insight into the fast pace of this class of passive, wave-based gates, relative to mechanical computing mechanisms which rely on quasi-static deformations, electroactive buckling, or snap-through soft elastomerics (typically in the order of tens of seconds) \citep{el2024intelligent, meng2021bistability}. The computing speed of the topology-optimized gates relies primarily on wave propagation speed and the transient interval each gate takes for the required output ratio to take shape. Time stamps reported here indicate that waves transmit through the gates within roughly $100$ microseconds, and achieve desired functionality less than half a millisecond from the moment of excitation.

\begin{figure*}[h!]
\centering
\includegraphics[width=\linewidth]{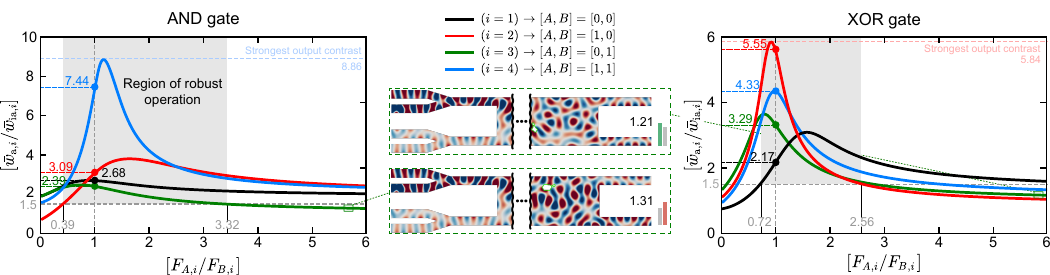}
\caption{Variations in the displacement output ratio, $[\bar{w}_{\mathrm{a},i}/\bar{w}_{\mathrm{ia},i}]$, as a result of changing the input excitation force ratio, $[F_{A,i}/F_{B,i}]$, across all four input combinations for an AND gate (left) and an XOR gate (right). In both gates, the ratio $[F_{A,i}/F_{B,i}]$ is swept from $0.01$ to $6$, and black, red, green, and blue curves correspond to $i=1$, $2$, $3$, and $4$, respectively. The two shaded regions denote the range of non-uniform forcing amplitudes within which all output ratios remain above the preset threshold of $1.5$, marking the robust operational region for each gate. For the AND gate, this region corresponds to $0.39 < [F_{A,i}/F_{B,i}] < 3.32$, with a strongest output ratio of $8.86$ attained in the ($i=4$) case. The XOR gate exhibits a slightly narrower region corresponding to $0.72 < [F_{A,i}/F_{B,i}] < 2.56$, with a strongest output ratio of $5.84$ attained in the ($i=2$) case. Intersections of each curve with the vertical $[F_{A,i}/F_{B,i}] = 1$ line represent the nominal operating points reported in Figure~\ref{Fig4}, corresponding to gate excitations of identical amplitudes. Outside the region of acceptable performance, at least one case in each gate produces an output ratio below the $1.5$ threshold, as demonstrated by the example wavefields in the middle insets, indicating that the gate no longer meets the required logical performance. The results demonstrate the compatibility of the optimized logic gates to uneven and unpredictable excitation amplitudes, which is commensurate with practical conditions.}
\label{Fig6}
\end{figure*}

\subsection{Sensitivity to non-identical input amplitudes and gate robustness}
\label{robustness}

The mechanical gates investigated in this work demonstrate a consistent performance in realizing the desired logical operation solely through their passive designs, when subject to two inputs of equal vibrational amplitudes. In other words, regardless of whether an input leg hosts an active state of (0) or (1), the simulations conducted in Figure~\ref{Fig4} impose both excitations with an equal forcing. In integrated circuits in which these individual gates serve as building blocks toward a larger operation, feeding an output from one gate as input to another, this condition of equal amplitudes should not be assumed. This is further confirmed by the wide range of output ratios, $[\bar{w}_{\mathrm{a},i}/\bar{w}_{\mathrm{ia},i}]$, displayed in the results, ranging from $2.39$ to $7.44$ for the AND gate and from $2.17$ to $5.55$ for the XOR gate. To address this, we conduct a robustness study to evaluate the efficacy of both gates when subject to inputs of non-identical vibrational amplitudes, and assess their sensitivity to such amplitude discrepancies.

The robustness study is conducted by varying the input excitation force ratio, $[F_{A,i}/F_{B,i}]$, where $F_{A,i}$ is the force applied at input $A$, and $F_{B,i}$ is the force applied at input $B$, for a given input scenario, $i$ (following the same definition of $i$ used earlier and indicated in Figure~\ref{Fig3}(b)). In this analysis, the ratio $[F_{A,i}/F_{B,i}]$ is varied from $0.01$ to $6$, evaluating the performance of both gates under a wide range of input force ratios. The resulting wavefields and output ratios $[\bar{w}_{\mathrm{a},i}/\bar{w}_{\mathrm{ia},i}]$ are analyzed to determine the range of input force ratios over which the gates maintain correct logical functionality. The results, illustrated in Figure~\ref{Fig6}, reveal that each gate exhibits a robust operating window, within which all logical states remain distinguishable with output ratios $[\bar{w}_{\mathrm{a},i}/\bar{w}_{\mathrm{ia},i}]>1.5$. The AND gate maintains proper functionality within the input force ratio $[F_{A,i}/F_{B,i}]$ range of $0.39$ to $3.32$, defining a safe operation region where the gate consistently performs as expected. A similar robustness assessment for the XOR gate indicates a narrower yet stable safe operation region ranging between $0.72$ and $2.56$. It should be noted that certain cases may still produce acceptable logical behavior well outside this ``safe'' region. However, the output ratio of at least one case falls below the required threshold, i.e., $[\bar{w}_{\mathrm{a},i}/\bar{w}_{\mathrm{ia},i}]<1.5$, outside of the shaded window, thus failing to meet the design criteria (see the two example wavefields at the center of Figure~\ref{Fig6}). This illustrated robustness of the gates is crucial for practical implementation, as real input sources, such as piezoelectric exciters, seldom possess identical excitation forces, particularly at ultrasound frequencies. Furthermore, this guarantees reliability when these gates are incorporated into more complex circuits.

\section{Experimental realization}
\label{experimental_setup}


\begin{figure*}[htbp!]
\centering
\includegraphics[width=\linewidth]{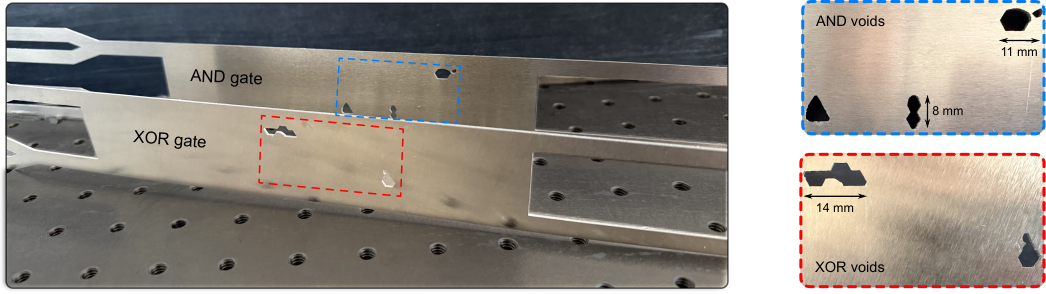}
\caption{Fabricated AND and XOR mechanical logic gates showing optimized void geometries and characteristic length scales.}
\label{Fig7}
\end{figure*}


\begin{figure*}[htbp!]
\centering
\includegraphics[width=\linewidth]{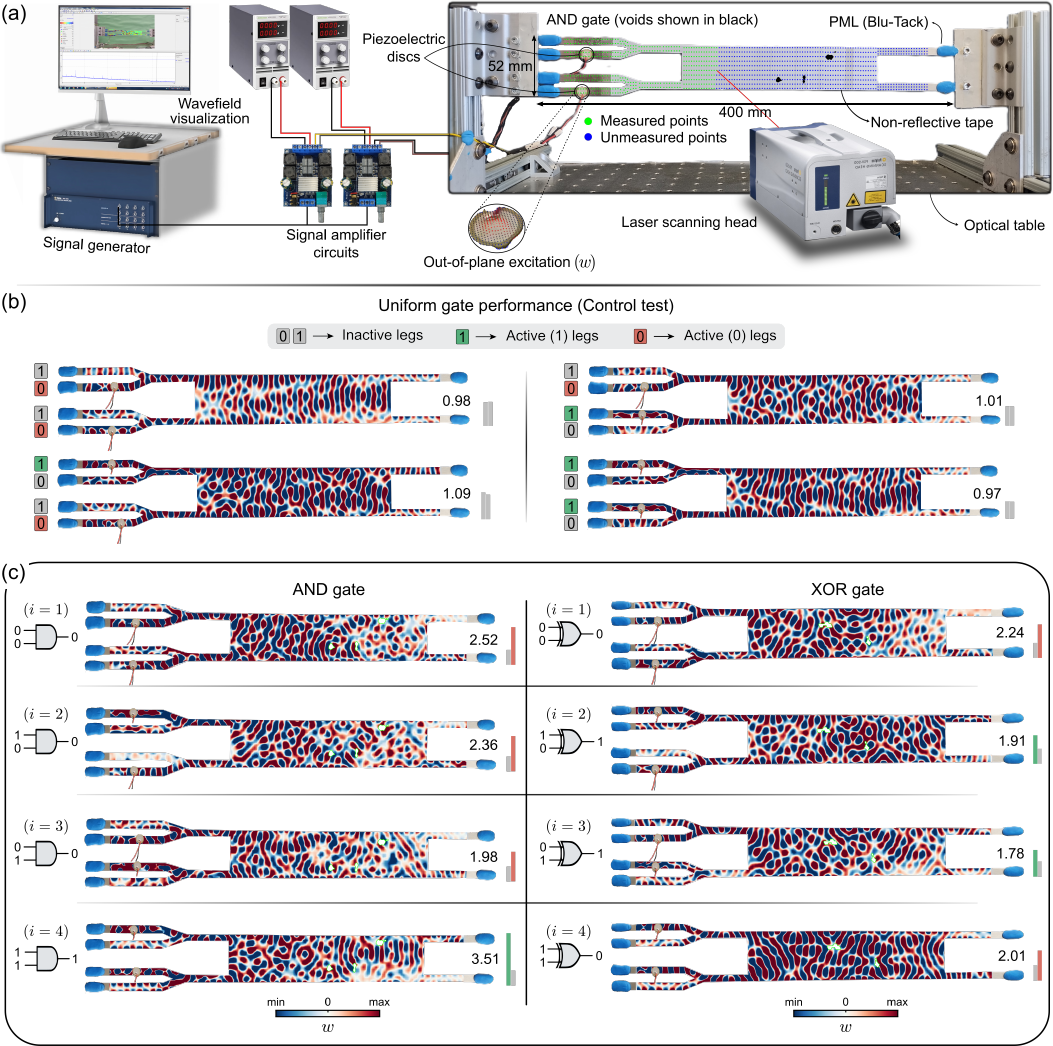}
\caption{Experimental realization of topology-optimized mechanical logic gates. (a) Detailed view of the experimental setup, showing an AND gate mounted between two clamps, with the laser scanning head measuring out-of-plane displacements over the indicated grid. The two piezoelectric discs are driven via signals generated by the vibrometer's front-end unit and excited in an out-of-plane mode, as shown in the inset. Each disc is independently amplified using separate amplifier circuits to ensure identical inputs to the gate. The gate, fabricated from a laser-cut aluminum sheet, is coated with non-reflective tape to enhance laser signal strength, and Blu-Tack adhesive is attached to the input and output legs to suppress reflections. (b) Experimental results for the control test case, i.e., a uniform gate with no voids in the design domain, serving as a baseline to provide a sanity check (equal outputs in both legs) and facilitate the subsequent assessment of the optimized gates. The gate exhibits similar behavior across all four cases with nearly identical outputs, demonstrating no logical behavior, as anticipated. (c) Experimental out-of-plane displacement wavefields for the AND and XOR gates, reconstructed from the scanned high-density grid points, capturing the intricate details of wave scattering through the gate and around the optimization-induced voids (highlighted by green boundaries). The results confirm the effectiveness of the topology-optimized gates, fabricated using a low-cost laser-cutting process, in realizing the intended logic operation, achieving output ratios ranging from $[\bar{w}_{\mathrm{a},i}/\bar{w}_{\mathrm{ia},i}] = 1.98$ to $3.51$, and from $1.78$ to $2.24$ in the AND and XOR gates, respectively.}
\label{Fig8}
\end{figure*}

To experimentally validate the performance of the topology-optimized logic gates, we present an experimental implementation of the mechanical configurations developed using the presented theoretical framework. Experimental validation is focused on reproducing the predicted logical behavior and output-ratio contrast of individual gates, rather than achieving pointwise agreement of wavefields. As shown in Figure~\ref{Fig7}, the experimental prototypes are fabricated from a $1$-mm thick aluminum plate (6061 aluminum alloy) via laser-cut manufacturing, which maintains the sharp boundaries and intricate details of the solid-void regions envisioned numerically. Each gate spans a length of $400$ mm with a maximum height of $52$ mm at the input side. In every gate, piezoelectric discs ($10$-mm uxcell piezoelectric transducers) are glue-bonded to the two active input legs, and the boundaries of the input and output legs are coated with a damping material (Blu-Tack adhesive) to emulate the perfectly matched layer in the numerical simulations. For each set of inputs in the truth tables, $i$, the two piezoelectric discs corresponding to the active input legs harmonically excite the structure at the operational frequency of $92$ kHz. The reference signal to the piezoelectric actuators is first generated by the front-end unit of a Polytec PSV-500 SLDV system and then fed to two custom-built circuits to separately amplify the signal strengths, powered by Eventek KPS3010D DC power supply, as shown in Figure~\ref{Fig8}(a). This setup allows us to independently control and fine-tune the signals driving each of the piezoelectric discs to account for any manufacturing discrepancies between the different discs, which may lead to non-identical excitation amplitudes (We note here that the gates maintain adequate performance even when subject to non-identical input forces, as demonstrated in Figure~\ref{Fig6}). As the elastoacoustic vibrational waves propagate through the different gates, the out-of-plane displacement wavefield, $w$, is captured using the SLDV over a scanning area with a resolution of $10$ points per wavelength. The reference excitation signals are fed back to the vibrometer to ensure measurement synchronization across the entire scanned region, allowing the collection of both amplitude and phase information at each point. 

We start the experimental study by testing the uniform gate, which represents the initial all-solid (aluminum) structure prior to optimization. This serves as a control test and a sanity check intended to confirm that the wavefield of a non-optimized gate maintains a symmetric pattern, showing no bias or significant wave energy difference between the two output legs, and practically failing to distinguish between an active and an inactive leg, i.e., $\bar{w}_{\mathrm{a},i} \approx \bar{w}_{\mathrm{ia},i}$. Indeed, the average displacement amplitudes within both output legs of the uniform gate prove to be nearly identical for all four input scenarios, as depicted in Figure~\ref{Fig8}(b). These results reinforce the notion that intelligence emerges only as a result of the optimized topology enabling the structure to execute the desired logical operation. 

Following the control test, the two optimized structures corresponding to AND and XOR gates were fabricated and experimentally measured, with all the ensuing wavefields shown in Figure~\ref{Fig8}(c). The results confirm that each gate reproduces the anticipated logical outcome under different input scenarios, with experimental output ratios, $[\bar{w}_{\mathrm{a},i}/\bar{w}_{\mathrm{ia},i}]$, that exceed the preset threshold of $1.5$ across the board. Specifically, the AND gate exhibits output ratios ranging from $1.98$ to $3.51$, while the XOR gate's corresponding ratios range from $1.78$ to $2.24$. The higher the ratio, the clearer the contrast is between the two output legs, promoting a strong performance by the gate, which is manifested by a better readability of the operation's outcome. As can be seen in Figure~\ref{Fig8}(c), this is best achieved by the ($i=4$) case of the AND gate, corresponding to $[A,B]=[1,1]$, and the ($i=1$) case of the XOR gate, corresponding to $[A,B]=[0,0]$. While minor discrepancies were observed between experimental and numerical results, particularly in the wavefield pattern and output ratios, these variations are practically unavoidable due to parametric uncertainties inherent in physical experiments. These include fabrication tolerances, geometric imperfections, and the difficulty of imposing input excitations of identical amplitudes to perfectly match the numerical simulations. Nevertheless, despite such inevitable input variations, the overall performance of both gates across all input configurations remains accurate, consistently generating the correct computational outcome for all tested cases. This robustness emphasizes the effectiveness of the TO algorithm in yielding complex mechanical structures capable of performing reliable mechanical logic, even in the presence of experimental uncertainties.

\begin{figure*}[h!]
\centering
\includegraphics[width=\linewidth]{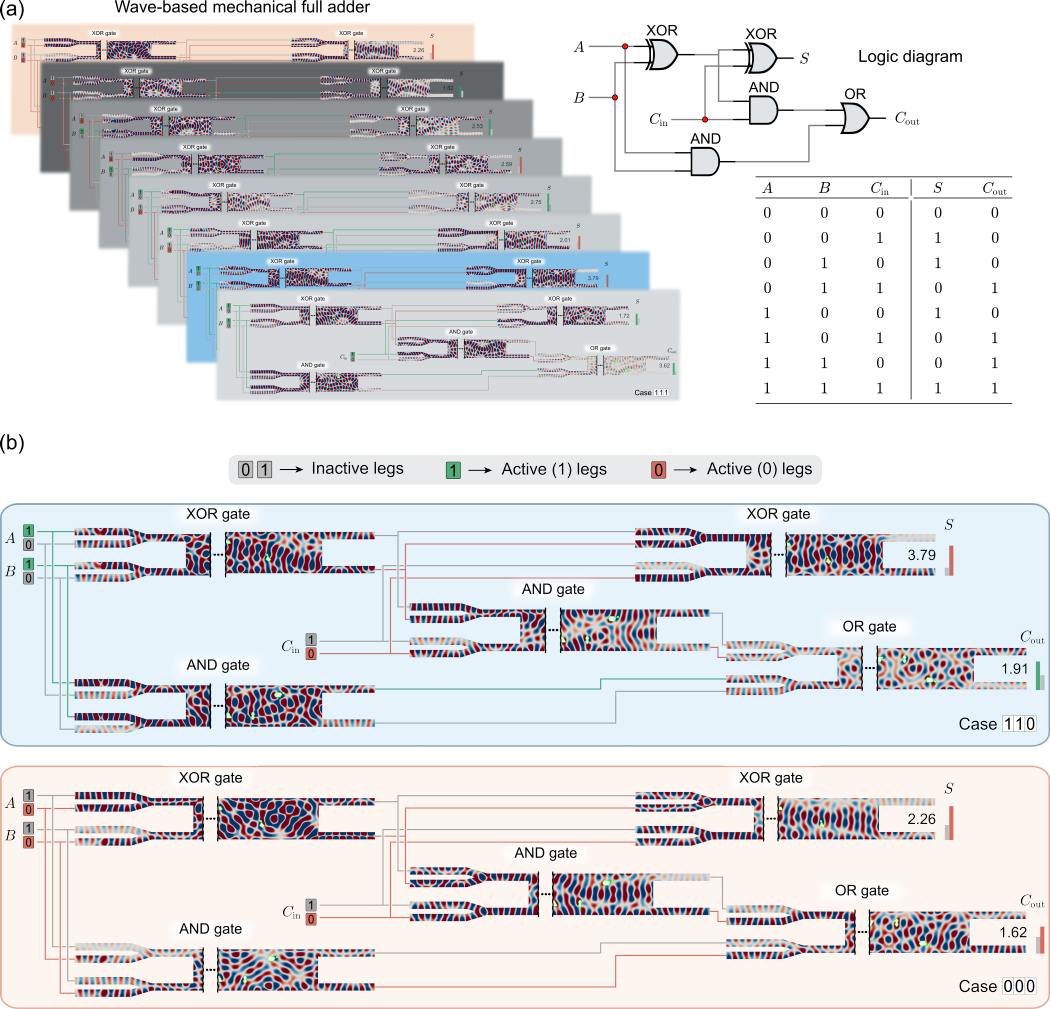}
\caption{Realization of a wave-based mechanical full adder circuit. (a) Different scenarios of a one-bit full adder, schematically illustrated using a logic diagram, are stacked in an order following the truth table shown on the right. The adder receives three binary inputs, $A$, $B$, and $C_{\mathrm{in}}$, yielding eight distinct configurations with outputs $S$ and $C_{\mathrm{out}}$. (b) Two select cases with high and low inputs are shown corresponding to the $[A, B, C_{\mathrm{in}}]=[1, 1, 0]$ (top) and $[A, B, C_{\mathrm{in}}]=[0, 0, 0]$ (bottom) cases, respectively. Connections between gates are indicated with solid lines: Green for an active (1) state, red for an active (0) state, and gray for inactive leg connections. Both cases demonstrate robust performance, consistently following the full adder truth table while maintaining output ratios $[\bar{w}_{\mathrm{a}}/\bar{w}_{\mathrm{ia}}]>1.5$, producing $S$ outputs of $3.79$ and $2.26$, and $C_{\mathrm{out}}$ outputs of $1.91$ and $1.62$, respectively, thus meeting the preset threshold, despite being fed non-uniform input amplitudes throughout the process.}
\label{Fig9}
\end{figure*}
\section{Mechanical full adder circuit}
\label{full_adder}

Logic gates are what allow computers to `think' and process data. Inside the central processing unit of computers, these gates interconnect to form circuits such as adders, multiplexers, and decoders that do math, make decisions, and transport data \citep{rafiquzzaman2005fundamentals, braun2014digital}. For instance, a full adder circuit comprises a network of two XOR gates, two AND gates, and one OR gate to add three one-bit binary numbers: Two inputs, $A$ and $B$, and a carry-in, $C_{\mathrm{in}}$, and produce a two-bit output: A sum, $S$, and a carry-out, $C_{\mathrm{out}}$. In other words, a single full adder can perform the addition of one-bit operands along with a carried input. By cascading 1-bit full adders, 4-bit, 8-bit, and 32-bit adders can be assembled, which form the basis of the ripple-carry adder \citep{ying2020electronic, vijay2022review, tari2019design}. The transition from a basic full adder circuit to a sophisticated, high-performance arithmetic network capable of executing extensive numerical operations with minimal latency represents a significant evolutionary advancement. In digital computing, accumulators, counters, and digital signal processors employ adder networks for ongoing numerical computations, whereas multipliers utilize extensive arrays of adders to consolidate partial products. Given such significance and utility, the creation of a wave-based adder circuit establishes the foundation for higher-order operations in passive mechanical form.

In this section, we present an integrated wave-based full adder circuit embodying the previously designed logic gates. Figure~\ref{Fig9}(a) shows the mechanical configuration and corresponding logic diagram of the adder, illustrating the basic operational concept. The overlaid snippets shown in the left panel of Figure~\ref{Fig9}(a) depict the eight possible input combinations of $A$, $B$, and $C_{\mathrm{in}}$ that the adder can be subject to, with their outcomes shown in tabular form (on the right side of the figure) and by virtue of the active output legs. All the shown cases utilize the same operating frequency as the fundamental gates. A full adder circuit has three consecutive layers of logic gates. The first layer comprises an XOR gate and an AND gate, which both receive the inputs $A$ and $B$. This is followed by a middle layer which contains an additional XOR gate and an additional AND gate that together receive inputs from the initial XOR gate as well as the carry-in value, $C_{\mathrm{in}}$. The readout of this second XOR gate directly represents the adder's sum, $S$. Finally, a third layer comprising a single OR gate receives inputs from both AND gates and outputs the adder's carry-out value, $C_{\mathrm{out}}$ (The reader is referred to Appendix~\ref{AND–to–OR}, detailing the re-purposing of an AND gate to yield OR operations through vertical inversion). All connections between layers are modeled using the continuity feature of the FE solver, allowing gates to share their output wavefields while preserving shape, amplitude, and phase information. These connections are color-coded in the figure, with green lines denoting active (1) states and red lines denoting active (0) states. All gray lines represent inactive states, where minimal forward-path vibrational energy flows.

With all the possible scenarios showcased in the upper panel of the figure, Figure~\ref{Fig9}(b) provides a closer look at two example cases: $[A, B, C_{\mathrm{in}}]=[1, 1, 0]$ (top) and $[A, B, C_{\mathrm{in}}]=[0, 0, 0]$ (bottom), with the goal of showing how the circuit behaves with low and high input combinations. In the $[1, 1, 0]$ case, the upper legs in both input ports are activated by incident waves in the XOR and AND gates within the first layer. Following their corresponding truth tables, the two gates generate outputs of (0) and (1), respectively. The second layer of XOR and AND gates is then activated by the first XOR gate output of (0) in addition to a $C_{\mathrm{in}}$ of (0). As a result, the output of the second XOR gate yields a sum, $S$, of (0), as a result of $0 \oplus 0 = 0$. Concurrently, the output of the second AND gate also yields (0), as a result of $0 \land 0 = 0$. Finally, the outputs of the two AND gates from the first and middle layers, (1) and (0), respectively, subsequently activate the OR gate in the final layer, yielding a $C_{\mathrm{out}}$ of (1), as a result of $1 \lor 0 = 1$. The mechanical full adder circuit conducts the $[1, 1, 0]$ case with output ratios of $3.79$ and $1.91$ for the $S$ and $C_{\mathrm{out}}$ outputs, respectively, well above the preset threshold of $1.5$. Similarly, the same circuit successfully executes the $[0, 0, 0]$ case with output ratios of $2.26$ and $1.62$, confirming the circuit's ability to effectively perform across various input combinations.

Going beyond the effective performance of the wave-based adder, the results shown in Figure~\ref{Fig9} truly demonstrate the versatility of the individual logic gates, when integrated as building blocks into a highly-interlaced circuit. The different output ratios of the individual gates, combined with the different network connections, trigger a series of non-identical input amplitudes between subsequent layers and backscattering within and between the interfacing gates as the calculation propagates downstream. For example, a close inspection of the two example cases shown in Figure~\ref{Fig9}(b) reveals that the incident waves impinging on the first layer first get mildly attenuated due to structural damping, resulting in non-uniform outputs which are passed forward to the second layer. However, as demonstrated in Figure~\ref{Fig6}, the gates are robust enough to maintain the desired logical performance across a wide range of input excitation ratios. Apart from this, incident waves received by the OR gate show more dissipation than their counterparts, having traveled through multiple layers of gates. This explains why, in both the $[1, 1, 0]$ and $[0, 0, 0]$ cases, the OR gate, as measured by its output ratio, underperforms that of the middle layer XOR gate, which generates the $S$ output. This effect could potentially be mitigated by employing a buffer circuit to restore the signal strength as needed. Secondly, minor backscattering appears throughout the structure, with a significant portion of these reflections corresponding to those previously observed in the inactive legs of the logic gates in Figure~\ref{Fig4}. Now in full adder configuration, these reflections can propagate through the preceding layers, giving the illusion that the gates fail to maintain the expected logical behavior. This wave leakage into the inactive legs varies between cases and appears to be more pronounced in the XOR and AND gates of the middle layer, as can be observed in Figure~\ref{Fig9}(b). Although backscattering control can improve stability, it risks compromising the optimization performance, and the phenomenon does not affect the intended performance of the forward path in any of the eight full adder cases.

\begin{table}[h!]
    \centering
    \caption{Output displacement ratios, $[\bar{w}_{\mathrm{a}}/\bar{w}_{\mathrm{ia}}]$, at the sum, $S$, and carry-out, $C_{\mathrm{out}}$, of the full adder circuit.}
    \label{tab1}
    \begin{tabular}{ccc|cc}
        \toprule
        \multicolumn{3}{c|}{{Input}} & \multicolumn{2}{c}{{Output ratio}} \\
        \cmidrule(lr){1-3} \cmidrule(lr){4-5}
        {$A$} & {$B$} & ${C_{\mathrm{in}}}$ &{$S$} & ${C_{\mathrm{out}}}$ \\
        \midrule
        0 & 0 & 0 & {2.26} & {1.62} \\
        0 & 0 & 1 & {1.62} & {1.92} \\
        0 & 1 & 0 & {2.53} & {1.67} \\
        0 & 1 & 1 & {2.59} & {3.15} \\
        1 & 0 & 0 & {2.75} & {2.00} \\
        1 & 0 & 1 & {2.01} & {2.04} \\
        1 & 1 & 0 & {3.79} & {1.91} \\
        1 & 1 & 1 & {1.72} & {3.62} \\
        \bottomrule
    \end{tabular}
\end{table}

Finally, Table~\ref{tab1} summarizes the output ratios, [$\bar{w}_{\mathrm{a}} / \bar{w}_{\mathrm{ia}}$], of the XOR and OR gates that generate the $S$ and $C_{\mathrm{out}}$ outcomes of the full adder obtained from numerical simulations, respectively, for all eight possible input combinations. These results confirm that the wave-based mechanical full adder is capable of accurately reproducing the truth table of its digital counterpart. Moreover, such performance consolidates the idea that complex circuits can be constructed by stacking mechanical logic gates, facilitating the scalable deployment of more sophisticated topology-optimized mechanical computing circuits.

\section{Concluding remarks}
\label{conclusion}

This study presents a robust and effective methodology for realizing passive, wave-based mechanical logic. The approach capitalizes on the rich dynamics of wave propagation and guidance in elastic substrates. Contrary to periodic and highly-ordered architected materials (e.g., metamaterials), we employ a radically different paradigm in which a topology optimization framework enables new configurations to be tailored for mechanical logic by satisfying preset objectives. Specifically, Boolean logic is executed via engineered scattering of vibroacoustic waves around strategically-placed voids in an elastic waveguide, which comprises input and output layers. While the shape, size, and location of such voids may not necessarily conform to human intuition, this symmetry-agnostic approach allows unconventional designs to be identified from an extremely large parameter space. Numerical studies demonstrate the successful design of topology-optimized logic elements ranging from fundamental gates, including AND, XOR, and OR, to a foundational combinational circuit of the full adder. These numerical findings are supported by carefully designed experiments, providing direct physical validation of the proposed mechanical logic structures. 

In the numerical studies, we employ a density-based TO algorithm that optimizes the distribution of an artificial material density variable, acting as the primary control parameter and directly governing the local stiffness and density fields. The objective of this optimization problem is the precise control of wave propagation, allowing constructive or destructive interference patterns of the scattered wavefield to guide vibrations toward designated output locations, thereby achieving the desired logical outcome. The results demonstrate the robust performance achieved across a variety of gates, in addition to the ability of the gates to adapt to binary inputs of varying and non-identical amplitudes. These outcomes are then experimentally validated through the fabrication and testing of the topology-optimized mechanical gates using a dedicated setup designed to evaluate their computational performance through laser-scanned wavefields. We further demonstrate the design scalability by integrating individual gates into a functional full adder circuit, highlighting the potential for assembling more complex architectures and advancing toward sequential wave-based logic. It is evident that, as gates are wave-based components, altering the excitation frequency changes the wavelength, and consequently, the wavefield pattern. As a result, mechanical wave-based intelligence is inevitably going to be sensitive to the excitation frequency (See Appendix~\ref{freqband} for a brief discussion of frequency bandwidth). Broadband wave-based logic can presumably be achieved by adopting a multi-frequency formulation, in which the objective function and ratio constraints are enforced over a set of frequencies covering the desired bandwidth. As expected, this approach involves a trade-off in the form of reduced peak contrast at the center frequency. Addressing and optimizing this trade-off can be the focus of future work.

Overall, this work establishes a flexible and extensible framework for the design of robust mechanical logic gates using a blank-slate, optimization-driven methodology. Compared to intelligent structures that rely on shape morphing or deformation, this wave-based approach to mechanical computing offers several advantages. Most importantly, the system presented here operates passively and attains swift computational speeds comparable to the underlying wave propagation velocities. Moreover, the versatility of the computing approach extends it to different modes and length scales other than those implemented in this work. These qualities render it an ultra-low-power candidate for embedded in-material computing, especially in applications where information has to be encoded mechanically without the need for traditional electronics. In environments where computational inputs are acoustic (speech/audio) or structural (vibrations), this form of mechanical intelligence supports low-latency, robust, and transduction-free sensing and information processing, and an ability to operate without electromagnetic interference. More broadly, these results highlight the potential of wave-based approaches for realizing complex, efficient, and perturbation-immune physical computing systems.

\section*{ORCID iDs}
\noindent Ali Jafari: \url{https://orcid.org/0000-0003-2523-6231}\\
Mohamed Mousa: \url{https://orcid.org/0000-0002-3561-9407}\\
Mostafa Nouh: \url{https://orcid.org/0000-0002-2135-5391}

\section*{Funding}
The authors disclosed receipt of the following financial support for the research, authorship, and/or publication of this article: This work was supported by the Mechanical Behavior of Materials program of the U.S. Army Research Office (ARO), under Grant No. W911NF-23-1-0078.

\section*{Declaration of conflicting interests}
The authors declare no potential conflicts of interest with respect to the research, authorship, and/or publication of this article.

\section*{Data availability statement}
The data supporting the findings of this study are available from the corresponding author upon reasonable request. All relevant data generated or analyzed during this study have been included in the article.



\appendix
\setcounter{section}{0}
\renewcommand{\thesection}{\arabic{section}}

\renewcommand{\thefigure}{\arabic{figure}}
\renewcommand{\theHfigure}{\arabic{figure}}

\newpage

\refstepcounter{section} 
\section*{Appendix 1} 
\vspace{-0.5em}\noindent{\textit{Gate optimization and performance under different initializations}}
\label{non-convex}

\begin{figure*}[h!]
\centering
\includegraphics[width=\linewidth]{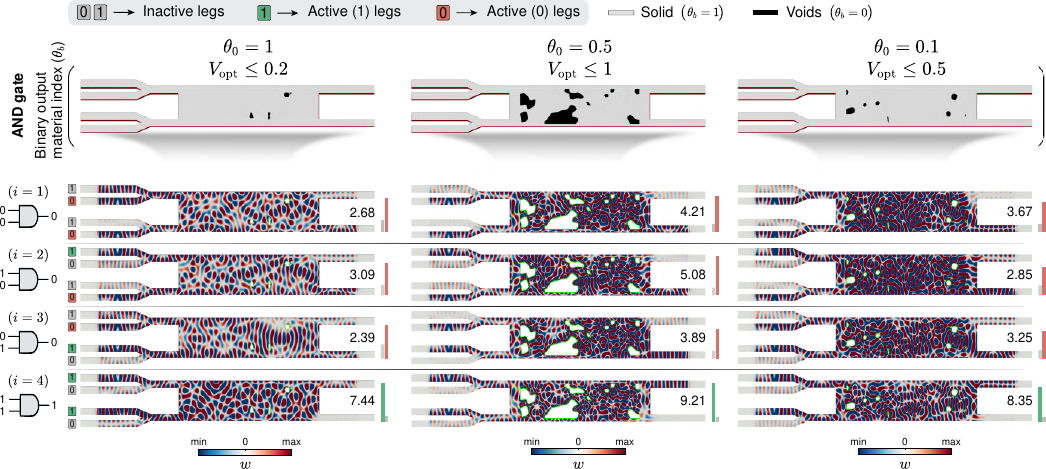}
\caption{Effect of different initial conditions on the optimized AND gate. The top row shows the final binary output material index for three uniform initializations ($\theta_0 = 1$, $0.5$, and $0.1$) under different volume constraints. The lower panels present the corresponding out-of-plane displacement fields for all four input combinations, along with the resulting output displacement ratios. While different initializations lead to distinct layouts, all designs correctly reproduce the intended logical outcome.}
\label{FigApp1}
\end{figure*}

Topology optimization is generally a non-convex problem which is sensitive to initial conditions. To assess sensitivity to initialization, additional numerical simulations were performed for the AND gate using different initial conditions applied to the design domain. As shown in Figure~\ref{FigApp1}, different initializations lead to distinct void patterns corresponding to different local optima. Nevertheless, all resulting designs achieve the required output contrast and correctly reproduce the intended truth-table behavior. The first column of Figure~\ref{FigApp1} shows the same AND gate presented earlier and experimentally investigated. In this design, we adopt a fully solid design initialization, $\theta_0=1$, as a standard and unbiased starting point for the design variables. Two other initializations corresponding to $\theta_0=0.5$ and $\theta_0=0.1$ are also provided. These initializations represent purely numerical starting guesses in which the optimization begins from an intermediate effective material density and subsequently increases or decreases the local density to converge towards a binary design. 

For the case with a fully solid initialization $\theta_0=1$, a volume constraint is imposed to prevent the optimized voids from occupying more than $20\%$ of the design domain ($V_{\mathrm{opt}}\leq 0.2$), thereby ensuring a bounded level of material removal consistent with the baseline gate design. For alternative initializations with intermediate values ($\theta_0=0.5$ and $\theta_0=0.1$), the optimization is intentionally allowed greater freedom in material redistribution, corresponding to increased admissible void volume fractions. In particular, the $\theta_0=0.5$ initialization permits void formation over the entire design domain ($V_{\mathrm{opt}}\leq1$), while the $\theta_0=0.1$ case allows up to approximately half the design-domain volume ($V_{\mathrm{opt}}\leq0.5$) to be removed. These variations are introduced to assess the robustness of the logical functionality under different initial conditions and volume constraints.

It is also important to note that granting greater freedom in the optimization domain can lead to stronger output contrast, reflecting a common tradeoff in topology-optimization problems. At the same time, increased material redistribution tends to produce more complex geometries, which can result in more intricate and nonlinear wave-manipulation behavior. Despite this increased complexity, the intended logical operation is preserved across all cases considered here, highlighting the robustness of the proposed wave-based logic framework.

\refstepcounter{section}
\section*{Appendix 2}
\vspace{-0.5em}\noindent{\textit{AND to OR transformation}}

\label{AND–to–OR}

The symmetric layout of the gate configuration along the $x$-axis, specifically, the mirrored positioning of the legs of the two inputs and the output, allows dual functionality between gate pairs that behave identically when the inputs match, as in cases ($i=1$) and ($i=4$), and oppositely when the inputs differ, as in cases ($i=2$) and ($i=3$). This symmetry pairs gates such as AND and OR gates, as well as their inverted variants. As a result, a single topology-optimized geometry can be utilized for both operations, allowing a primary readout (e.g., AND) and a secondary one which can be obtained through simple geometric transformation, i.e., vertical flipping (e.g., OR). This significantly reduces optimization cost, manufacturing effort, and overall footprints, while also improving fabrication consistency and enabling interchangeable use of the resulting gates in practical implementations.

\begin{figure*}[h!]
\centering
\includegraphics[width=\linewidth]{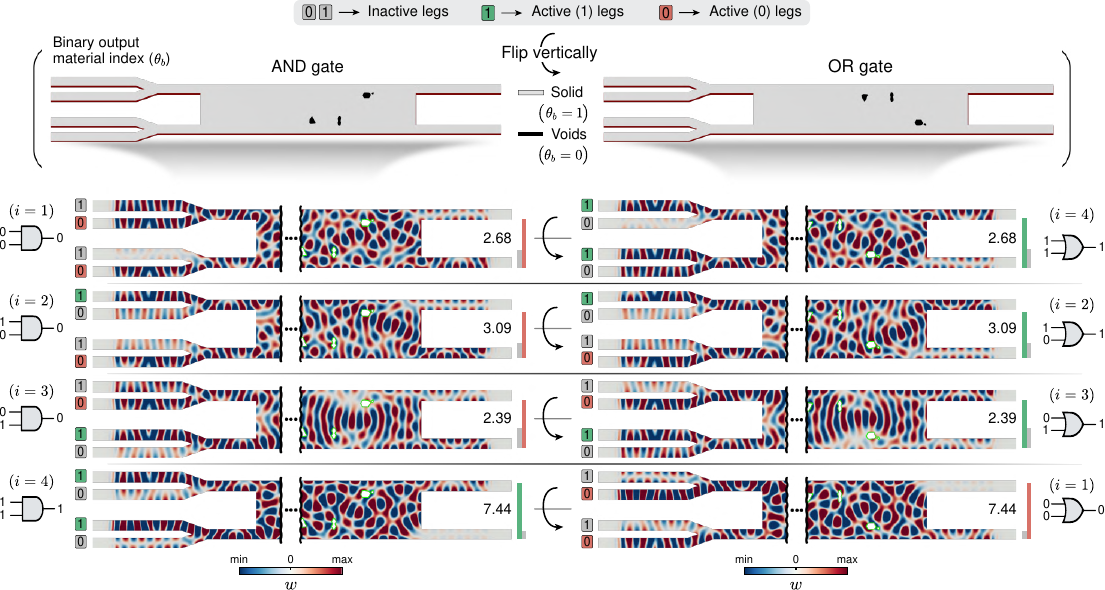}
\caption{A mechanical OR logic gate derived from the vertical inversion of a successful AND gate configuration. The vertical flip exchanges the (0) and (1) input terminals, resulting in the ($i=1$) and ($i=4$) cases being interchanged while the ($i=2$) and ($i=3$) remain unchanged, thereby achieving full OR functionality within the same optimized geometry.}
\label{FigApp2}
\end{figure*}

To elaborate more on the mechanism behind this dual functionality, we re-establish the labeling convention used throughout the study, with the upper legs of the input ports, $A$ and $B$, representing a (1) state, and the lower legs denoting a (0) state. By vertically flipping the AND gate and redefining the labeling such that the upper and lower legs correspond to (0) and (1) states, respectively, cases ($i=1$) and ($i=4$) become effectively swapped, while cases ($i=2$) and ($i=3$) remain unaltered, matching the expected truth table of an OR gate, as illustrated in Figure~\ref{FigApp2}. For instance, in case ($i=1$) of the AND gate, both inputs and the output are in state (0). After vertical inversion, these states become (1), corresponding to case ($i=4$) of the OR gate. In case ($i=2$) of the AND gate, i.e., $[A,B] = [1,0]$ with an outcome of (0), flipping the geometry doesn't affect the input configuration (inputs positions and states are both switched) but results in a (1) outcome, consistent with the OR gate behavior of case ($i=2$), as depicted in the second row of Figure~\ref{FigApp2} wavefield results.

\newpage

\refstepcounter{section}
\section*{Appendix 3}
\vspace{-0.5em}\noindent{\textit{Frequency bandwidth and performance robustness}}
\label{freqband}

The operation of the proposed logic gates relies on elastic-wave interference, making their performance inherently frequency dependent. Variations in the excitation frequency, therefore, modify the underlying wavefield patterns and can influence logical performance. In here, we present a brief frequency sensitivity analysis of the AND gate as a representative example to assess the bandwidth of robust gate operation. The analysis is performed by evaluating the output displacement ratios over a range of excitation frequencies around the design frequency, $92$ kHz, for all four input combinations of the AND gate. For a given threshold value $r$, a frequency is considered operationally robust if the logical classification of all input cases remains unchanged, i.e., if the output ratios consistently satisfy the truth-table requirements.

\begin{figure*}[h!]
\centering
\includegraphics{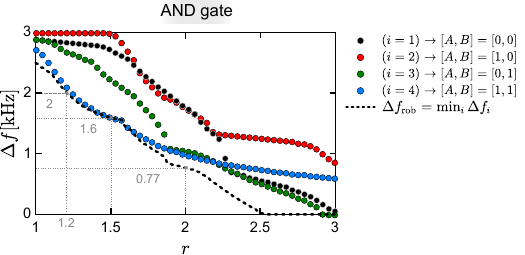}
\caption{Frequency robustness of the AND gate as a function of the threshold ratio $r$. The robust operational bandwidth $\Delta f$ is shown for each input combination (colored markers), while the dashed curve denotes the intersection of all four cases and defines the overall frequency window in which correct logical classification is preserved. At the nominal threshold $r=1.5$, the AND gate sustains a robust bandwidth of $1.6$ kHz centered at $92$ kHz; decreasing the threshold to $r=1.2$ increases the bandwidth to $2$ kHz, while increasing the threshold to $r=2$ reduces it to $0.77$ kHz.}
\label{FigApp3}
\end{figure*}

Figure~\ref{FigApp3} shows the resulting robust frequency window $\Delta f$ as a function of the threshold ratio $r$. The colored markers correspond to the robustness limits associated with individual input combinations, while the dashed curve corresponds to the overall robust bandwidth, defined as $\Delta f_{\mathrm{rob}}=\min_{i} \Delta f_{i}$, i.e., the frequency window over which all truth-table cases of the AND gate remain valid. As expected, increasing the threshold $r$ imposes a stricter separation between logical states, which enhances noise tolerance but reduces the admissible frequency window. For the threshold value used throughout the main manuscript, $r=1.5$, the AND gate exhibits a robust operational bandwidth of approximately $1.6$ kHz centered at the design frequency of $92$ kHz. Decreasing the threshold to $r=1.2$ increases the bandwidth to approximately $2$ kHz, while increasing the threshold to $r=2$ reduces it to approximately $0.77$ kHz, illustrating the tradeoff between logical contrast and frequency robustness. These results demonstrate that the gates possess a finite but well-defined operational bandwidth and that the threshold parameter $r$ provides a tunable means to balance robustness against frequency sensitivity. Similar behavior is expected for the other logic gates presented in this work, as they rely on the same wave-interference-based operating principles.

\end{document}